\documentclass[10pt,twocolumn,letterpaper]{article}

\usepackage[accsupp]{axessibility} 

\usepackage{ijcb}
\usepackage{times}
\usepackage{epsfig}
\usepackage{amssymb}

\usepackage{enumitem}
\usepackage{microtype}
\usepackage{booktabs} 
\usepackage{latexsym}
\RequirePackage{fix-cm}
\usepackage{amsmath, mathtools}
\usepackage{relsize}
\usepackage{footnote}
\usepackage[font=small,skip=0pt]{caption}
\usepackage[rightcaption]{sidecap}
\makesavenoteenv{tabular}
\makesavenoteenv{table}
\usepackage{multirow}
\usepackage{array, url}
\usepackage{dblfloatfix}
\usepackage{graphicx}
\DeclareGraphicsExtensions{.pdf,.jpg,.png}
\graphicspath{{figures/}}
\usepackage{wrapfig}
\usepackage{longtable}
\usepackage{lscape}
\usepackage{pifont}

\newcolumntype{L}[1]{>{\raggedright\let\newline\\\arraybackslash\hspace{0pt}}m{#1}}
\newcolumntype{C}[1]{>{\centering\let\newline\\\arraybackslash\hspace{0pt}}m{#1}}
\newcolumntype{R}[1]{>{\raggedleft\let\newline\\\arraybackslash\hspace{0pt}}m{#1}}

\newcommand*{\thead}[1]{
\multicolumn{1}{c}{\bfseries\begin{tabular}{@{}c@{}}#1\end{tabular}}}

\usepackage[pagebackref=true,breaklinks=true,letterpaper=true,colorlinks,bookmarks=false]{hyperref}

\ijcbfinalcopy 

\ifijcbfinal\fi
\begin{document}

\title{Benchmark Dataset Dynamics, Bias and Privacy Challenges in Voice Biometrics Research}

\author{Casandra Rusti
\thanks{This study was done while the authors collaborated on the Fair EVA project (\href{https://faireva.org}{https://faireva.org})}\\
\normalsize University of Southern California\\
\normalsize United States\\
{\tt\footnotesize rusti@usc.edu}
\and
Anna Leschanowsky\footnotemark[1]\\
\normalsize Fraunhofer IIS\\
\normalsize Germany\\
{\tt\footnotesize leschaaa@iis.fraunhofer.de}
\and
Carolyn Quinlan\footnotemark[1]\\
\normalsize University of Toronto\\
\normalsize Canada\\
{\tt\footnotesize carolyn.quinlan@mail.utoronto.ca}
\and
Michaela Pnacek(ova)\footnotemark[1]\\
\normalsize York University\\
\normalsize Canada\\
{\tt\footnotesize pnaci@yorku.ca}
\and
Lauriane Gorce\footnotemark[1]\\
\normalsize Open North\\
\normalsize Canada\\
{\tt\footnotesize lauriane@opennorth.ca}
\and
Wiebke (Toussaint) Hutiri\footnotemark[1]\\
\normalsize Delft University of Technology\\
\normalsize The Netherlands\\
{\tt\footnotesize w.toussaint@tudelft.nl}
}


\maketitle
\thispagestyle{empty}

\begin{abstract}
Speaker recognition is a widely used voice-based biometric technology with applications in various industries, including banking, education, recruitment, immigration, law enforcement, healthcare, and well-being. However, while dataset evaluations and audits have improved data practices in face recognition and other computer vision tasks, the data practices in speaker recognition have gone largely unquestioned. Our research aims to address this gap by exploring how dataset usage has evolved over time and what implications this has on bias, fairness and privacy in speaker recognition systems. Previous studies have demonstrated the presence of historical, representation, and measurement biases in popular speaker recognition benchmarks. In this paper, we present a longitudinal study of speaker recognition datasets used for training and evaluation from 2012 to 2021. We survey close to 700 papers to investigate community adoption of datasets and changes in usage over a crucial time period where speaker recognition approaches transitioned to the widespread adoption of deep neural networks. Our study identifies the most commonly used datasets in the field, examines their usage patterns, and assesses their attributes that affect bias, fairness, and other ethical concerns. Our findings suggest areas for further research on the ethics and fairness of speaker recognition technology.
\end{abstract}


 \section{Introduction}
\label{introduction}


Speaker recognition is widely used in voice biometrics in the private and public sectors, e.g, to verify the identity of banking clients~\cite{TDPersonalBankingVoicePolicy, BMOVoiceWorks} or employees~\cite{ThomasHeath2016ThisBathroom}, and to secure an expanding network of voice assistants and voice-based Internet of Things devices through which people interact with digital services~\cite{Seaborn2021voice}. The large scale deployment of speaker recognition systems has been facilitated by the adoption of deep learning approaches that have greatly improved technology performance~\cite{Bai2021Speaker}. However, using data-intensive, deep learning systems in the development of biometric systems can lead to bias and discrimination, characterized by disparate error rates across demographic groups~\cite{Cavazos2021AccuracyBias, Krishnapriya2019characterizing, DeFreitasPereira2022fairness, Hutiri2022BiasRecognition}. Nonetheless, despite rapid progress and widespread adoption of speaker recognition technology, bias, discrimination and fairness remain largely unexplored in voice biometrics.  

The biometrics community has noted the importance of evaluating bias. For example, the NIST Face Recognition Vendor Test now includes an evaluation of bias across demographic groups~\cite{NIST2021Eval}, and several studies have evaluated error rate disparities across groups in face recognition models~\cite{Cavazos2021AccuracyBias, DeFreitasPereira2022fairness, Kotwal}. Model bias however presents only one of several sources of bias in the development of deep learning systems~\cite{Suresh2021Framework} and training and evaluation datasets should be additionally interrogated~\cite{Paullada2021DataResearch}. Bias in training datasets reflects downstream in the learned models, while bias in evaluation datasets skews evaluation outcomes, channels future development efforts and makes it impossible to assess if models are biased~\cite{Suresh2021Framework}. Motivated by prior research on dataset evaluations~\cite{Raji2021AboutEvaluation, Hirota2022GenderDatasets, Pahl2022FemaleFaces} and development~\cite{Sambasivan2021EveryoneAi, Liao2021AreLearning}, this paper presents the first study on the impact of training and evaluation datasets on bias and other ethical concerns in the voice biometrics domain.



This study aims to explore dataset usage dynamics in order to gain insights into community adoption of datasets and potential cultural shifts in data practices that are likely to impact the development of speaker recognition technology. We address the following questions to examine how the adoption of deep learning has shifted data practices and to extrapolate how data and evaluation practices may impact fairness and privacy concerns in voice biometrics: 
\begin{enumerate}[nosep]
    \item Which datasets are used for training and evaluation in speaker recognition research? 
    \item How has dataset usage changed over the period from 2012 to 2021? 
    \item What are the attributes of the most used datasets?
    \item What are the implications of the above questions for bias, fairness, privacy and other ethical challenges in speaker recognition?
\end{enumerate}

While prior dataset studies have been published in speaker recognition~\cite{Campbell1999CorporaSystems, Sturim2016CorporaSystems}, they focused primarily on describing available corpora and did not interrogate how data practices impact ethical and societal outcomes. This paper presents insights on development practices in the speaker recognition domain, and contributes to the emerging body of work on bias, fairness and privacy challenges in biometrics. We start by reviewing 
related literature in Section~\ref{s:related_work} before describing our research approach in Section~\ref{s:method}, and presenting results in Section~\ref{s:results}. We consolidate our findings and reflect on the study in Section~\ref{s:discussion}. 

\section{Related Work}
\label{s:related_work}
The section covers prior research on bias in face recognition systems, measures for evaluating demographic bias in biometric verification systems, and the impact of bias in data and datasets on machine learning models. 

\subsection{Bias in Biometrics}
Existing bias literature on biometrics mainly focuses on measuring disparate error rates across demographic groups in face recognition systems. Studies have proposed various measures, such as statistical methods and Fairness Discrepancy Rate, for assessing the demographic bias of biometric verification systems ~\cite{Kotwal, DeFreitasPereira2022fairness}. Meanwhile, others have provided checklists for measuring race bias in face recognition emphasizing 
the need to consider data-driven factors and scenario modelling including 
accounting for sub-population distributions, algorithm quality, the representation of and conditions captured by images, threshold selection and appropriate considerations around demographic pairing~\cite{Cavazos2021AccuracyBias}. 

In the voice biometrics domain, Hutiri and Ding present an empirical and analytical examination of bias in the machine learning development workflow of speaker verification benchmarks~\cite{Hutiri2022BiasRecognition}. They identify historical, representation and measurement bias during the data gathering stage, and learning, evaluation, aggregation and deployment bias during model building. In particular, pairing of trials in speaker recognition benchmarks can result in evaluation datasets of variable difficulty across demographic groups~\cite{Hutiri2022DesignDatasets} showing similarities to bias evaluation for face recognition~\cite{Cavazos2021AccuracyBias}. 

\subsection{Bias in Data and Datasets}
Machine learning models are impacted by bias in data and datasets, including historical, representation, measurement, and evaluation bias~\cite{Sambasivan2021EveryoneAi, Suresh2021Framework}. As datasets form the basis for training, evaluating, and benchmarking models, dataset evaluations are important to interrogate bias in machine learning systems~\cite{Paullada2021DataResearch}. Prior research has found that the dominant developer culture, which emphasizes rapid progress and ever-larger models, can lead to representation bias in datasets and inadequate dataset documentation~\cite{Paullada2021DataResearch}. Similarly, evaluation failures can result from implementation variations, errors in test set construction, overfitting, and inadequate baselines~\cite{Liao2021AreLearning}.


This study takes inspiration from recent dataset evaluations in the visual domain~\cite{Raji2021AboutEvaluation, Pahl2022FemaleFaces, Hirota2022GenderDatasets}. We particularly draw on the work of Raji and Fried, who studied the detailed make-up of evaluation datasets and benchmarks used for face recognition~\cite{Raji2021AboutEvaluation}. They surveyed over 100 face datasets to consider the selection of evaluation tasks, benchmark data, evaluation criteria and metrics. Their approach revealed insights into the historical evolution of face recognition, driven by evaluation challenges of the NIST, with funding from intelligence agencies. The advancement of speaker recognition research shares many common attributes with that of face recognition. In contrast to Raji and Fried's study, we however examine dataset usage in the speaker recognition research community over a period of time.

\section{Research Approach}
\label{s:method}

\begin{figure}[hbt]
    \centering
    \includegraphics[width=0.9\linewidth]{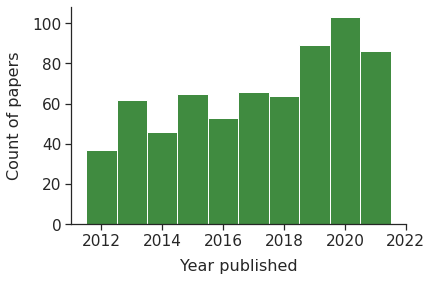}
    \caption{Histogram of papers published at Interspeech over the past decade and analyzed in this study.}
    \label{fig:interspeech_paper_count}
\end{figure}

Next, we describe our approach to studying the dynamics of speaker recognition dataset adoption and use in the research literature over the past decade. The study focuses on peer-reviewed research published over a ten year period from 2012 - 2021 at the Interspeech conference, one of two main international conference venues for academic and industrial speech research. We included all papers from the International Speech Communication Association (ISCA) archive\footnote{\url{https://www.isca-speech.org/archive/}} that contained the search terms \textit{speaker recognition}, \textit{speaker verification}, \textit{speaker identification} or \textit{speaker authentication} in their title or abstract. This query resulted in 696 papers, which we analyzed further. We excluded 25 papers that were overview papers, studied speaker recognition by humans or applications rather than model development, or did not explicitly mention which datasets were used to train and evaluate their models. Our final analysis thus includes 671 papers. Over the decade that we analysed, the number of papers on speaker recognition published at Interspeech has doubled, as can be seen in Figure~\ref{fig:interspeech_paper_count}.

All papers were tagged with the training and evaluation datasets that they used. As datasets were not always named consistently, some assumptions were made. For example, studies that used datasets from the NIST Speaker Recognition Evaluations (SREs) - regularly and typically annually released for benchmarking purposes - often rely on several of these datasets for training and refer to them as a range (e.g. NIST 2004 - 2008). 
In these cases, we assumed that every dataset in the range was used for training or evaluation, as indicated by the authors. Overall, we encountered many naming inconsistencies. Especially evaluation datasets were sometimes only referred to on the highest level (e.g. VoxCeleb), without specifying which dataset version, subset or evaluation protocol was used. Whenever possible, we standardized dataset and subset names and otherwise tagged papers by their training and evaluation dataset \textit{family}.
The dataset family name was created by manually cleaning the dataset names, and then extracting the first word in the name as the family name. It is common in the speech domain to refer to training datasets as development data or corpora, and to evaluation datasets as test data. In this paper, we use the terms training and evaluation datasets, unless we refer to the names of specific datasets.

\section{Speaker Recognition Dataset Dynamics over a Decade of Use} \label{s:results}

This section examines which datasets the speaker recognition research community has adopted over the past decade, how the use of datasets for training and evaluation purposes has changed over time, and which attributes characterize the most frequently used datasets. This provides insight into data practices and usage dynamics and their implications on bias, fairness and ethics in the speaker recognition domain. 

\subsection{Community Adoption of Speaker Recognition Datasets} \par \label{ss:dataset_adoption} 

\begin{figure*}[hbt]
    \centering
    \includegraphics[width=\textwidth]{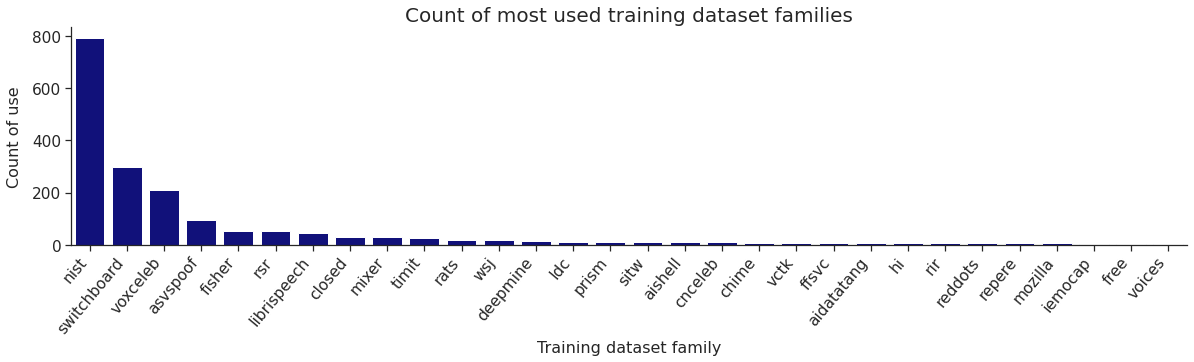}
    \caption{Histogram showing the count of most used training dataset families between 2012 and 2021.}
\label{fig:train_frequency}
\end{figure*}

Over the past decade, a wide range of datasets has been used for training and evaluating speaker recognition systems. In total, we counted 185 unique training and 164 unique evaluation dataset families. Despite this variety, a small number of dataset families has dominated speaker recognition research. Figure~\ref{fig:train_frequency} shows the frequency counts of the top 30 dataset families used for training and evaluation respectively in the analyzed papers. As papers could use more than one dataset to train and evaluate systems, the total count of dataset families exceeds the number of analyzed papers. 

The NIST Speaker Recognition Evaluation (SRE) corpora dominate both training and evaluation. These corpora were not unique datasets in their own right, but rather collections and subsets of other datasets, predominantly Switchboard and Mixer. We have kept the NIST SRE labels distinct from Switchboard and Mixer to stay true to the naming conventions used by researchers. Moreover, the NIST SREs typically required specific settings for training and evaluation that did not necessarily include the entire datasets. The NIST SREs were both users and drivers of these dataset collections, as annual evaluation challenges required new datasets to evaluate speaker recognition technology in ever more difficult settings~\cite{Cieri2007}. Table~\ref{tab:top10training} shows that Mixer and Switchboard datasets are commonly used, in the NIST SREs, among the top ten training datasets.  

\begin{table}[hbt]
    \centering
    \begin{tabular}{l|c}
    \textbf{Dataset} & \textbf{times used} \\ \midrule
        NIST SRE 04 (Switchboard, Mixer) & 136\\
        NIST SRE 05 (Mixer) & 133\\
        NIST SRE 06 (Mixer) & 127\\
        NIST SRE 08 (Mixer) & 96\\
        VoxCeleb 2 & 76\\
        VoxCeleb 1 & 52\\
        Switchboard (version not specified) & 48\\
        Switchboard Cellular 2 & 43\\
        VoxCeleb 1 - dev & 43\\
        NIST SRE 10 (Switchboard, Mixer, etc.) & 40\\
    \end{tabular}
    \caption{Top 10 most frequently used training datasets. \protect\footnotemark}
    \label{tab:top10training}
\end{table}

After the NIST SREs, Switchboard occurs second most frequently for training, but surprisingly is only rarely used for evaluation. We elaborate on this later in Section~\ref{ss:datasetattributes}. 
After the NIST SRE corpora, VoxCeleb and ASVspoof have been popular datasets for training and evaluation. ASVspoof are datasets released by the Automatic Speaker Verification and Spoofing Countermeasures Challenge\footnote{\href{https://www.asvspoof.org/}{https://www.asvspoof.org/}}, which was launched in 2015 to address growing concerns of security breeches in speaker verification technology due to voice spoofing and deepfakes. In later sections, we elaborate on VoxCeleb and its rise to popularity. 



\begin{figure}[t]
    \centering
    \includegraphics[width=0.9 \linewidth]{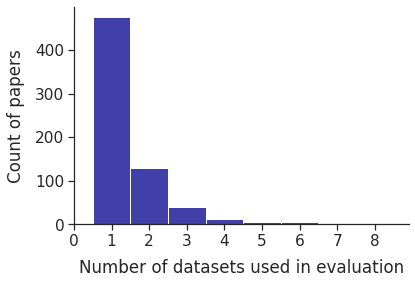}
    \captionof{figure}{Histogram of number of datasets used for evaluation by each paper that we reviewed. Most papers use a single dataset.}
    \label{fig:hist_evaldataset_count}
\end{figure}

Overall, papers used a far greater number of datasets for training than for evaluation. Figure~\ref{fig:hist_evaldataset_count} shows that the majority of papers in our study used only a single dataset for evaluation. Papers that evaluated on more datasets rarely used more than three. These numbers are low, and reminiscent of evaluation failures in machine learning more broadly~\cite{Liao2021AreLearning}. While speaker recognition development on limited corpora that target specific and evolving tasks over an extended period of time may have been justified to advance the field prior to the adoption of deep neural networks~\cite{Sturim2016CorporaSystems}, the same practices today will lead to overfitting. Consequently, we ought to question the validity of claimed advances.


\subsection{Dataset Dynamics over a Decade of Use} \par
\label{ss:dataset_dynamics}

\begin{figure*}[hbt]
    \centering
    \includegraphics[width=\textwidth]{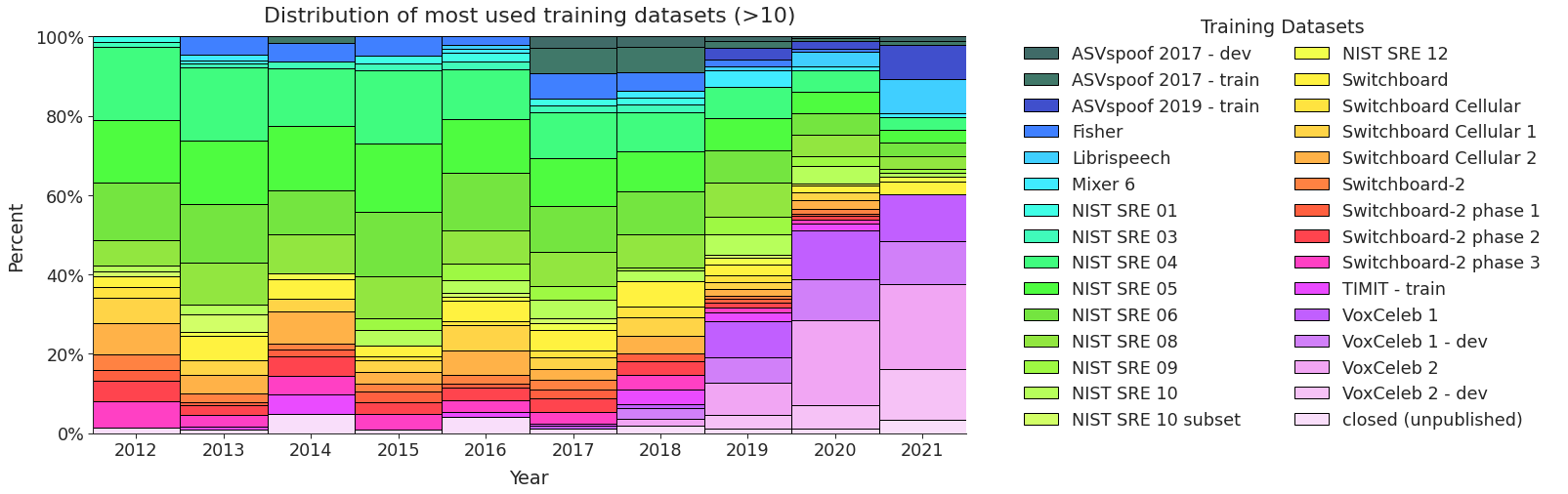}
    \caption{Distribution (\%) of dataset use for speaker recognition \textbf{training} (VoxCeleb datasets purple, NIST SREs turquoise \& green).}
\label{fig:train_dataset_density}
\end{figure*}

\begin{figure*}[hbt]
    \centering
    \includegraphics[width=\textwidth]{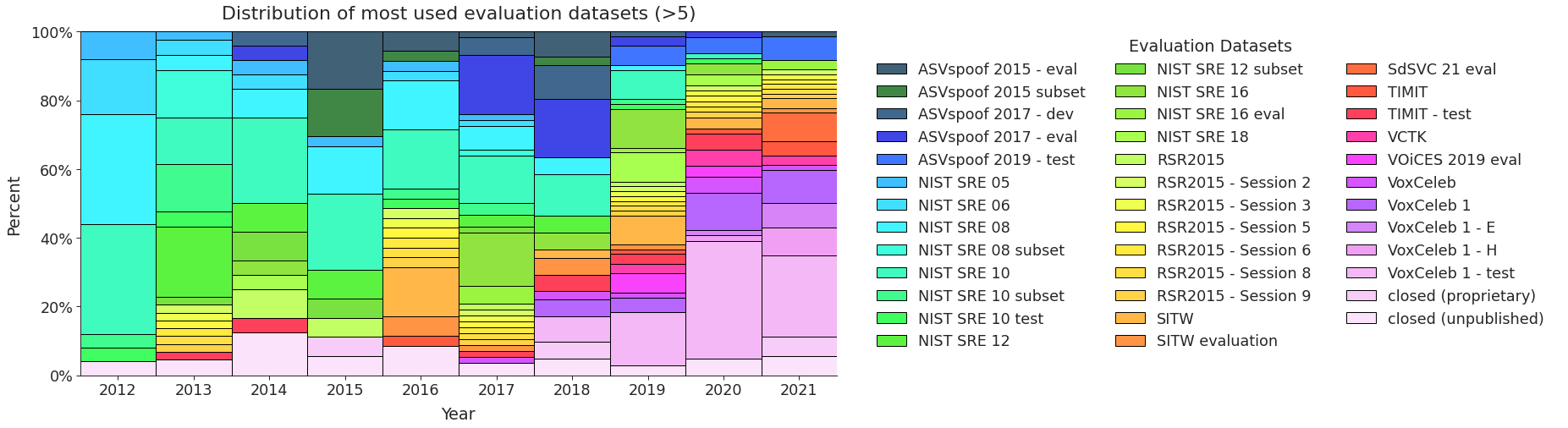}
    \caption{Distribution (\%) of dataset use for speaker recognition \textbf{evaluation} (VoxCeleb datasets purple, NIST SREs turquoise \& green).}
\label{fig:eval_dataset_density}
\end{figure*}

Next, we examine changes in dataset usage, and by extension community adoption of datasets over the past decade. Where the NIST SREs, Switchboard and Mixer datasets featured prominently in our previous analysis where we considered aggregate counts over the past decade, examining dataset usage year-on-year reveals that their popularity has declined dramatically. In their stead, VoxCeleb now dominates speaker recognition training and evaluation. Figures~\ref{fig:train_dataset_density} and~\ref{fig:eval_dataset_density} illustrate these dataset dynamics by visualizing the proportional use of datasets in training and evaluation. These figures show densities and should be considered together with Figure~\ref{fig:interspeech_paper_count}, which shows the growth in publications and consequently also dataset usage over the decade. Thus, since 2017 more papers have been published in speaker recognition, and a greater proportion of these studies uses VoxCeleb to train and evaluate their models.

Particularly striking is the extent to which VoxCeleb1 dominates speaker recognition evaluations. The dataset is disjoint from its successor, VoxCeleb2. A popular pairing is thus to use the larger VoxCeleb2 dataset for training, and VoxCeleb1 for evaluation. In 2020 and 2021 VoxCeleb1 made up over half of all evaluation dataset usage. More so, VoxCeleb1-test, a small subset of 40 predominantly male, US speakers whose name starts with an \textit{E}~\cite{Hutiri2022BiasRecognition} is used in a significant proportion of evaluations. As mentioned previously, papers may use more than one dataset for evaluation. We thus investigate the number of papers relying solely on VoxCeleb, or on its even more limited subset VoxCeleb1-test for evaluation in Table~\ref{tab:eval_with_voxceleb}. This additional analysis reveals that VoxCeleb1 is not only popular for evaluation, but that a significant proportion of studies evaluate their approaches on a single VoxCeleb dataset only (2019: 17\%, 2020: 21\%, 2021: 12\%), or even more restricted on VoxCeleb1-test only (2019: 12\%, 2020: 13\%, 2021: 8\%). 


\begin{table}[hbt]
\small 
\begin{tabular}{c|cccc}
\textbf{Year} & \thead{Published\\papers} & \thead{1 evaluation \\dataset only} & \thead{1 VoxCeleb \\dataset only} & \thead{VoxCeleb1 \\- test only} \\ \midrule
\textbf{2012} & 37 & 26 (70\%) & - & - \\
\textbf{2013} & 62 & 46 (74\%) & - & - \\
\textbf{2014} & 46 & 33 (72\%) & - & - \\
\textbf{2015} & 65 & 55 (85\%) & - & - \\
\textbf{2016} & 53 & 39 (74\%) & - & - \\
\textbf{2017} & 66 & 47 (71\%) & 1 & - \\
\textbf{2018} & 64 & 40 (63\%) & 4 (6\%) & 2 (3\%) \\
\textbf{2019} & 89 & 59 (66\%) & 15 (17\%) & 11 (12\%) \\
\textbf{2020} & 103 & 75 (73\%) & 22 (21\%) & 13 (13\%)  \\
\textbf{2021} & 86 & 55 (64\%) & 10 (12\%) & 7 (8\%)  \\
\end{tabular}
\caption{Most speaker recognition studies evaluate their models on a single dataset, with a large number only using one VoxCeleb dataset, or even just VoxCeleb1-test.}
\label{tab:eval_with_voxceleb}
\end{table}


\subsection{Dataset Attributes} \par
\label{ss:datasetattributes}

The past two sections have highlighted that the Switchboard, Mixer and VoxCeleb datasets, together with the NIST Speaker Recognition Evaluations (SRE) have dominated the development of speaker recognition technologies, each in its own era. Therefore, we now investigate attributes of these dataset families and their influence on bias, fairness and privacy of the technology today.

\medskip \noindent \textbf{Background and Motivation for Corpora Collection} \par 
\noindent The collection and release of the Switchboard corpora started in the 1990s and continued through the early 2000s. In total, seven datasets of two-sided English language telephone conversations were released. The dataset collections were funded by the US Defense Advanced Research Projects Agency (DARPA) and the US Department of Defense. The Linguistic Data Consortium (LDC) was primarily responsible for data collection and management. According to the LDC, these datasets were intended for "research, development, and evaluation of automatic systems for speech-to-text conversion, talker identification, language identification and speech signal detection purposes"~\cite{Graff2004SwitchboardCellularPart2}.

The Mixer and Transcript Reading (short Mixer) corpora succeeded the Switchboard corpora, as the collection protocol of the latter became complicated, time-consuming and expensive. Moreover, telephone behaviour of people changed as cellphones became popular~\cite{Cieri2007}. The Mixer project aimed to support various speaker recognition tasks in multi-lingual and cross-channel settings with channel referring to the medium used for speech recordings, e.g., microphone types. 
Mixer was created by the LDC in collaboration with the Lincoln Laboratory, the US NIST and the Speaker Identification research community.

The VoxCeleb datasets were a timely response to an increasing appetite in the speaker recognition community to test and develop their approaches in more challenging real-world (i.e. "in the wild") settings. VoxCeleb1 was released in 2017 by the Visual Geometry Group (VGG) at the University of Oxford, with the goal of creating a large scale, text-independent speaker recognition dataset that mimics unconstrained, real-world speech conditions \cite{Nagrani2017VoxCeleb}. A key driver for this was to explore the use of deep neural networks (DNNs), which had gained traction in computer vision, for speaker recognition tasks. A year later, VGG released VoxCeleb2 to expand the original data collection. 

\clearpage

\medskip \noindent \textbf{Attributes and Usage}\footnote{We did not purchase a membership or licenses Therefore, for the analysis of attributes we relied on openly accessible information provided by the LDC and papers on these corpora. We did not have access to the Switchboard and Mixer datasets for an in-depth analysis of attributes.} \par
\noindent Most of the Switchboard datasets contain more than 2~000 recordings with over 100 hours of speech. The Switchboard Credit Card dataset is an exception with only 35 recordings and 227 minutes of speech data. At the time of collection, these datasets were considered rich data sources for training speaker recognition models. Demographic metadata in Switchboard includes information on age, sex, years of completed education, country of birth, city and state where raised. The gender distribution in Switchboard is generally balanced on a speaker level, but not reported for the number of conversation samples. Switchboard Cellular Part~2 Audio reported gender demographics across recordings and has an overall balanced split in male and female representation across the dataset versions (with some years skewed 5-10\% towards one gender).  The recordings are annotated for channel quality (e.g. echo, crosstalk, static) and background noise. Switchboard Cellular Part~2 Audio also annotates the speaker's environment (indoors, outdoors, moving vehicle). The earlier Switchboard datasets record landline conversations with a variety of telephones, and the later datasets cellphone conversations. Including different handsets was always considered important to capture channel variability, which has a major influence on signal processing. Our analysis in Section~\ref{ss:dataset_dynamics} shows that Switchboard was predominantly used for training, with relatively limited use in evaluation. The reason for this is that speakers appear across multiple recordings in different dataset releases. However, it is not possible to connect speakers between the various releases. Using Switchboard for training and evaluation thus has potential for data leakage, which diminishes the quality of an evaluation. 

Mixer recorded significantly more data than Switchboard, with individual datasets capturing between 5~000 and 20~000 calls, resulting in tens of thousands of hours of speech. The early phases of the Mixer corpora focused on multi-lingual data collection. Focus later shifted towards multi-channel set-ups. Publicly available information for the Mixer corpora mostly contains details on the multi-channel set-up and recording devices. Descriptions of speaker metadata~\footnote{\href{https://catalog.ldc.upenn.edu/docs/LDC2020S03/readme.txt}{https://catalog.ldc.upenn.edu/docs/LDC2020S03/readme.txt} and \href{https://catalog.ldc.upenn.edu/docs/LDC2013S03/readme.txt}{https://catalog.ldc.upenn.edu/docs/LDC2013S03/readme.txt}} for Mixer 4, 5, and 6 give insights into the demographics that were collected which include sex, year of birth, education, occupation, ethnicity, height and weight, smoking and information about his or her family. Given this detailed metadata, the Mixer collection has been used for age estimation~\cite{Sadjadi2016}, smoker identification~\cite{Ma2022a, Ma2022b} and for predicting speaker demographics from word usage~\cite{Gillick2010}. These studies have found that the number of recordings in Mixer decreases drastically with age, and that non-smokers and female speakers are overly represented. Based on these observations, an extensive dataset evaluation of the Mixer corpora would be beneficial to identify sources of bias and their influence on the development of speech-related technology. 

The VoxCeleb datasets were scraped from celebrity YouTube videos to capture a large number of audio clips where people speak in unconstrained settings. VoxCeleb1 consists of 153~516 speech utterances from 1~251 speakers. VoxCeleb2 contains 1~128~246 utterances from 6~112 speakers. The creators of the dataset promote its use for speaker identification and verification, speech separation (i.e. diarization), talking face synthesis, cross-modal transfer between face and voice (i.e. making inferences about somebody's face based on their voice, and vice versa), emotion recognition and face generation. The only metadata available are gender and nationality labels. The dataset descriptions are not transparent about how gender labels were obtained, but it is likely that they came from VGGFace1 and 2~\cite{Parkhi2015Deep,Cao2018VGGFace2:Age}, which provided the candidate list of speakers to include in VoxCeleb. The nationality labels were inferred from speakers' countries of citizenship, as obtained from Wikipedia. The motivation for doing this was to assign a label that is indicative of a speaker's accent~\cite{Nagrani2020a}. The authors claim that the datasets are gender balanced, with 55\% and 61\% male speakers in VoxCeleb1 and 2 respectively. 

Perhaps the biggest difference between VoxCeleb and the other two datasets is that it used to be freely available for download\footnote{VGG has recently removed the public download link and added a privacy note to their website}. Switchboard and Mixer require a subscription to the LDC (\$3~850 for universities, \$27~500 for corporations) or must be purchased. Licensing costs for an individual dataset range between \$100 and \$300, however, not all datasets can be licensed without an LDC membership. This made the VoxCeleb datasets the first large scale, freely available datasets for speaker recognition. 



\medskip \noindent \textbf{Influence of Collection Method on Bias} \par
\noindent For Switchboard and Mixer participants were recruited to meet the language requirements of the study. They received financial compensation for participating (this changed over time, with start of the Mixer collection compensation changed from a fixed compensation per call to a per-call incentive and completion bonuses~\cite{CieriMixer1}). 
While the free phone call was a strong incentive to participate when the data collection started, after the turn of the millennium it lost its appeal~\cite{Cieri2007}. Between the releases of the Switchboard and Mixer datasets, the team of data collectors had continuity. This has led to consistency in the data collection method. However, it also implies that individual biases that shaped the design of the datasets can have gone unnoticed. For example, it is unclear how categories for labels such as ethnicity or "country where raised" were chosen, and what rules were applied to ensure consistent labelling. Such labeling choices can lead to measurement bias, which affects the validity of predictions made by speaker recognition systems. The target population for the Switchboard data collection were native English speakers in the American South. Most participants were college students, resulting in a dataset with a young demographic.

English had always been a dominant language in speaker recognition. The first three phases of the Mixer project thus focused on collecting multilingual data from bilingual speakers. 16\% of Mixer calls in Phases 1 and 2 are in Arabic, Mandarin, Russian or Spanish~\cite{CieriMixer}. The Mixer 3 collection also aimed at supporting language recognition, and had more than 2900 participants making calls in 19 different languages~\cite{Cieri2007}. The defense backing of the Mixer datasets is evident in the languages that were selected for the project, and their connection to US national security and military interests. The different Mixer corpora include conversations in Arabic, Egyptian, Farsi, Bengali, Hindi, Urdu, Tamil, 4 dialects of Chinese (also Mandarin), Japanese, Korean, Tagalog, Thai, Vietnamese, German, Italian, Russian, 3 dialects of English (including American  English), Spanish and Canadian French. Mixer 4, 5 and 6 featured a wider variety of channels and recording scenarios. The variety of languages and accents decreased as a consequence and the collections focused on native speakers of American English only~\cite{BrandscheinMixer4}. The data collectors attempted to balance dialects by recruiting 25\% of participants from Philadelphia, 25\% from Berkeley, and specifically from Texas, Georgia, Illinois, and New York~\cite{BrandscheinMixer4}. On-site recordings for Mixer 4 and 5 were carried out at two different locations, the LDC in Philadelphia, Pennsylvania, and at the International Computer Science Institute (ICSI) in Berkeley, California. Recruitment for Mixer 6 was done at the LDC~\cite{BrandscheinMixer4, BrandscheinMixer6}, thus decreasing the likelihood of collecting speech samples from speakers of various dialects and increasing the risk of representation bias.

The VoxCeleb datasets were constructed with a fully automated data processing pipeline from audio-visual media scraped from YouTube~\cite{Nagrani2017VoxCeleb, Nagrani2020a}. Both data pipelines consist of the same processing steps: first select a list of candidate speakers, then download videos from YouTube, apply face tracking, identify active speakers, verify identities from faces, remove duplicates, and finally find associated nationality metadata on Wikipedia. The candidate speakers for the datasets were sourced from VGGFace1~\cite{Parkhi2015Deep} and VGGFace2~\cite{Cao2018VGGFace2:Age} respectively. A previous study highlighted that this automated processing pipeline reinforces popularity bias from search results in candidate selection, and directly translates bias in facial recognition systems into the speaker recognition domain~\cite{Hutiri2022BiasRecognition}. Moreover, celebrities, especially actors and singers, have a high degree of control over their voice and accent, and should not be assumed to represent ordinary conversational speech.

\medskip \noindent \textbf{More than Bias: Privacy Threats and Ethical Questions} \par

\noindent The Switchboard and Mixer dataset collections passed an institutional ethical review\footnote{Guidelines of the Institutional Review Board of the University of Pennsylvania} and the LDC kept personal identifying information separate from the recordings~\cite{CieriMixer2, BrandscheinMixer4}. Nonetheless, the two corpora would today be considered as posing significant privacy risks to study participants. The privacy risks stem from two sources, firstly the content of the conversations and secondly the rich metadata, which makes it possible with today's data processing techniques to retrospectively correlate personal attributes with voice characteristics. The amount of personal information stored in the metadata is quite extensive and makes it possible to use these datasets for various tasks that include the identification of personal information from speech data in future. At the time of data collection, the ethical consequences and privacy concerns due to the extent of personal information contained in the voice may not have been clear to researchers. However, a decade of progress in speech science has changed that~\cite{Singh2019Profiling}. We are not aware that any efforts have been made to address the presence of personally identifiable information and sensitive attributes in the recordings, the potential impact on data subjects as well as the risks of re-identification. In today's data-driven society, privacy and anonymity of data subject are vital concerns that require attention and proper measures. 

During data collection, participants were asked to discuss a specific topic with an automated operator on a phone call, but to withhold personal information. Yet, the topics provided for discussion included political, cultural, social and religious topics\footnote{In the European Union the General Data Protection Regulation (GDPR) considers personal data revealing ethnic origin or religious beliefs as particularly sensitive and allows processing only on certain legal bases~\cite{GDPR2016}}. Another category of interviews took the form of informal conversations, adapted from sociolinguistic interview modules. Here subjects were encouraged to describe events of the past~\cite{BrandscheinMixer6}. This form of interview creates the illusion of an informal setting, making it more likely that participants share personal information with interviewers~\cite{Rodrigues2012}. While participants were not forced to discuss the topic provided, most of them followed the suggestion~\cite{BrandscheinMixer4}. Moreover, participants might have shared sensitive information during phone calls or interviews, increasing the risk of re-identification. For instance, Mixer 5 interviews covered family and personal history, raising the likelihood of participants shareing personal stories with similarly sensitive information as collected in the metadata. To our knowledge calls were not redacted to exclude personal information.  

The VoxCeleb datasets present very different privacy concerns. As has been the case with other web-scraped datasets, the dataset creators did not obtain consent from data subjects. Initially, they justified their approach by claiming that these datasets were "open-source" media. More recently, they have added a privacy notice to their website, calling on a data protection exemption of the University of Oxford based on Article 14(5)(b) of the UK GDPR, which allows data processing for scientific or historic purposes. Considering the highly sensitive nature of voice data, the military and security foundations of speaker recognition, and the wide-scale applications in the surveillance industry, it seems prudent to interrogate whether this exception ought to apply to voice data collected for speaker recognition purposes.

\section{Discussion}
\label{s:discussion}
The NIST SREs, Switchboard, and Mixer datasets have significantly influenced speaker recognition research over two decades. An important focus of this research has been on addressing audio processing challenges and reducing intra-speaker variability to ensure robustness of voice biometrics technology~\cite{Zheng2017robustness, Roux2014ADatasets, Sturim2016}. The NIST's evaluation-driven research agenda evolved alongside technology advancements, considering varied task environments, collection devices, background noise, and room acoustics. However, inter-speaker differences related to demographics or other speaker-dependent attributes are still considered secondary. Building robust systems to address inter-speaker variability is crucial to avoid biased and discriminatory systems from being deployed in critical applications, such as financial systems and voice-activated emergency response.


This study highlights how the shift to deep neural networks in speaker recognition has led to changes in research and data practices. Our analysis clearly illustrates the dominance of the VoxCeleb datasets for training and evaluation. These datasets met the demand of researchers to study speaker recognition systems with deep neural networks in unconstrained, "in-the-wild" settings. A subtle value-shift seems to prioritize studying challenging "in-the-wild" settings, potentially at the cost of developing systems that cater to diverse users. These shifts resemble observations made by~\cite{Paullada2021DataResearch} about data practices in machine learning research.

\subsection{Recommendations}

Speaker recognition technology offers benefits but also poses potential risks and harms depending on its deployment and use. Given recent advances in voice cloning~\cite{le2023voicebox}, anti-spoofing research should consider bias and fairness, to ensure that all demographic groups are adequately protected.  When used for voice-based authentication and access control, it is crucial to ensure the technology works for all users. To do this, representative evaluation datasets are necessary. 

To promote fairness and reduce bias in voice biometric datasets curation, valuable insights can be drawn from prior work in facial recognition~\cite{Raji2021AboutEvaluation}. Diverse and representative datasets that accurately reflect the demographics of the population being served are needed. Factors such as age, gender, race, and ethnicity should be considered when selecting people for dataset collection.  Furthermore, representation should be ensured at the speaker and utterance level to ensure equitable evaluation across demographic groups~\cite{Hutiri2022DesignDatasets}. Additionally, dataset collection procedures and dataset attributes should be documented carefully, for example by adopting datasheets~\cite{Gebru2021datasheets} for voice biometric datasets. More research is needed to understand application-specific requirements and how to incorporate these into evaluation protocols.

Beyond performance disparities, speaker recognition systems contribute to a hidden and pervasive surveillance infrastructure that enables governments and corporations to identify citizens and extract sensitive personal information from their voice. From a surveillance perspective, speaker recognition technology poses privacy risks to citizens. Striving for more representative datasets or detecting and mitigating bias can unintentionally increase harm to citizens rather than reduce it. Continued research efforts are needed to enable private and privacy-preserving voice processing. 

\subsection{Limitations}

We acknowledge limitations in our study and research approach. Our focus was on studying dataset dynamics, and we did not consider evaluation protocols and metrics, which are also important in speaker recognition evaluations. By analyzed peer-reviewed publications from the Interspeech conference, we assumed that dataset dynamics in the research domain are indicative of adoption and attitudes towards datasets in the voice biometrics community. We speculate that these research practices extend to industry and may indicate potential bias in deployed applications. Further studies and technology audits are necessary to evaluate bias in speaker recognition to enable accountability, transparency, and auditability in speaker recognition. 

\section{Conclusion}
\label{s:conclusion}
Our research provides a comprehensive overview of the evolution of speaker recognition datasets used for training and evaluation over the past decade. Through our analysis of the adoption, dynamics, and attributes of these datasets, we have identified potential issues related to bias, fairness, and ethical concerns in speaker recognition technology. These findings emphasize the importance of ongoing  investigation into dataset attributes and usage, particularly in light of current research practices. Moreover, our study highlights the need for further exploration of how data and evaluation practices may impact the ethics and fairness of speaker recognition technology. These insights contribute to our understanding of how data practices in this field may influence the future development of voice technologies, raising awareness about potential biases or fairness concerns.



\section{Acknowledgements} 
\noindent This research was supported by a Mozilla Foundation grant. 

{\small
\bibliographystyle{ieee}
\bibliography{references, manual_references}

\begin{thebibliography}{10}\itemsep=-1pt

\bibitem{GDPR2016}
{Regulation (EU) 2016/679 of the European Parliament and of the Council of 27
  April 2016 on the protection of natural persons with regard to the processing
  of personal data and on the free movement of such data, and repealing
  Directive 95/46/EC (General Data Protection Regulation) (Text with EEA
  relevance) BibSonomy}.

\bibitem{Bai2021Speaker}
Z.~Bai and X.~L. Zhang.
\newblock {Speaker recognition based on deep learning: An overview}.
\newblock {\em Neural Networks}, 140:65--99, 2021.

\bibitem{BMOVoiceWorks}
{BMO}.
\newblock {Voice recognition banking: See how our voice ID works}.

\bibitem{BrandscheinMixer4}
L.~Brandschain, C.~Cieri, D.~Graff, A.~Neely, and K.~Walker.
\newblock Speaker recognition: Building the mixer 4 and 5 corpora.

\bibitem{BrandscheinMixer6}
L.~Brandschain, D.~Graff, C.~Cieri, K.~Walker, C.~Caruso, and A.~Neely.
\newblock The mixer 6 corpus: Resources for cross-channel and text independent
  speaker recognition.

\bibitem{Campbell1999CorporaSystems}
J.~P. Campbell and D.~A. Reynolds.
\newblock {Corpora for the evaluation of speaker recognition systems}.
\newblock {\em ICASSP, IEEE International Conference on Acoustics, Speech and
  Signal Processing - Proceedings}, 2:829--832, 1999.

\bibitem{Cao2018VGGFace2:Age}
Q.~Cao, L.~Shen, W.~Xie, O.~M. Parkhi, and A.~Zisserman.
\newblock {VGGFace2: A dataset for recognising faces across pose and age}.
\newblock In {\em 13th IEEE International Conference and Workshops on Automatic
  Face and Gesture Recognition}, page 826, 2018.

\bibitem{Cavazos2021AccuracyBias}
J.~G. Cavazos, P.~J. Phillips, C.~D. Castillo, and A.~J. O'Toole.
\newblock {Accuracy Comparison across Face Recognition Algorithms: Where Are We
  on Measuring Race Bias?}
\newblock {\em IEEE Transactions on Biometrics, Behavior, and Identity
  Science}, 3(1):101--111, 1 2021.

\bibitem{CieriMixer}
C.~Cieri, W.~Andrews, J.~P. Campbell, G.~Doddington, J.~Godfrey, S.~Huang,
  M.~Liberman, A.~Martin, H.~Nakasone, M.~Przybocki, and K.~Walker.
\newblock The mixer and transcript reading corpora: Resources for multilingual,
  crosschannel speaker recognition research *.

\bibitem{CieriMixer2}
C.~Cieri, W.~Andrews, J.~P. Campbell, G.~Doddington, J.~Godfrey, S.~Huang,
  M.~Liberman, A.~Martin, H.~Nakasone, M.~Przybocki, and K.~Walker.
\newblock {The Mixer and Transcript Reading Corpora: Resources for
  Multilingual, Crosschannel Speaker Recognition Research *}.
\newblock Technical report.

\bibitem{CieriMixer1}
C.~Cieri, J.~P. Campbell, H.~Nakasone, D.~Miller, and K.~Walker.
\newblock {The Mixer Corpus of Multilingual, Multichannel Speaker Recognition
  Data * * * *}.
\newblock Technical report.

\bibitem{Cieri2007}
C.~Cieri, L.~Corson, D.~Graff, and K.~Walker.
\newblock Resources for new research directions in speaker recognition: The
  mixer 3, 4 and 5 corpora.
\newblock volume~4, pages 2864--2867, 2007.

\bibitem{DeFreitasPereira2022fairness}
T.~De~Freitas~Pereira and S.~Marcel.
\newblock {Fairness in Biometrics: A Figure of Merit to Assess Biometric
  Verification Systems}.
\newblock {\em IEEE Transactions on Biometrics, Behavior, and Identity
  Science}, 4(1):19--29, 2022.

\bibitem{Gebru2021datasheets}
T.~Gebru, J.~Morgenstern, B.~Vecchione, J.~W. Vaughan, H.~Wallach, H.~D. Iii,
  and K.~Crawford.
\newblock {Datasheets for datasets}.
\newblock {\em Communications of the ACM}, 64(12):86--92, 2021.

\bibitem{Gillick2010}
D.~Gillick.
\newblock Can conversational word usage be used to predict speaker
  demographics?
\newblock In {\em Interspeech 2010}, pages 1381--1384. ISCA, Sept. 2010.

\bibitem{Graff2004SwitchboardCellularPart2}
D.~Graff, K.~Walker, and D.~Miller.
\newblock Switchboard cellular part 2 audio, 2004.

\bibitem{Hirota2022GenderDatasets}
Y.~Hirota, Y.~Nakashima, and N.~Garcia.
\newblock {Gender and Racial Bias in Visual Question Answering Datasets}.
\newblock In {\em ACM Fairness Accountability and Transparency (FAccT) '22},
  2022.

\bibitem{Hutiri2022DesignDatasets}
W.~Hutiri, L.~Gorce, and A.~Y. Ding.
\newblock {Design Guidelines for Inclusive Speaker Verification Evaluation
  Datasets}.
\newblock In {\em Proc. Interspeech 2022}, pages 1293--1297, Incheon, Repbulic
  of Korea, 2022. International Speech Communication Association.

\bibitem{Hutiri2022BiasRecognition}
W.~T. Hutiri and A.~Y. Ding.
\newblock {Bias in Automated Speaker Recognition}.
\newblock In {\em ACM International Conference Proceeding Series}, pages
  230--247. Association for Computing Machinery, 6 2022.

\bibitem{Kotwal}
K.~Kotwal and S.~Marcel.
\newblock {Fairness Index Measures to Evaluate Bias in Biometric Recognition}.
\newblock pages 1--14.

\bibitem{Krishnapriya2019characterizing}
K.~S. Krishnapriya, K.~Vangara, M.~C. King, V.~Albiero, and K.~Bowyer.
\newblock {Characterizing the variability in face recognition accuracy relative
  to race}.
\newblock {\em IEEE Computer Society Conference on Computer Vision and Pattern
  Recognition Workshops}, 2019-June:2278--2285, 2019.

\bibitem{le2023voicebox}
M.~Le, A.~Vyas, B.~Shi, B.~K.~L. Sari, R.~Moritz, M.~Williamson, V.~Manohar,
  Y.~Adi, J.~Mahadeokar, and W.-N. Hsu.
\newblock Voicebox: Text-guided multilingual universal speech generation at
  scale.
\newblock 2023.

\bibitem{Liao2021AreLearning}
T.~I. Liao, R.~Taori, I.~D. Raji, and L.~Schmidt.
\newblock {Are We Learning Yet? A Meta-Review of Evaluation Failures Across
  Machine Learning}.
\newblock In {\em Neural Information Processing Systems (NeurIPS 2021)}, 2021.

\bibitem{Ma2022a}
Z.~Ma, Y.~Qiu, F.~Hou, R.~Wang, J.~T.~W. Chu, and C.~Bullen.
\newblock Determining the best acoustic features for smoker identification.
\newblock pages 8177--8181, 5 2022.

\bibitem{Ma2022b}
Z.~Ma, S.~Singh, Y.~Qiu, F.~Hou, R.~Wang, C.~Bullen, and J.~T.~W. Chu.
\newblock Automatic speech-based smoking status identification.
\newblock pages 193--203. Springer International Publishing, 2022.

\bibitem{Nagrani2020a}
A.~Nagrani, J.~S. Chung, W.~Xie, and A.~Zisserman.
\newblock {Voxceleb: Large-scale speaker verification in the wild}.
\newblock {\em Computer Speech and Language}, 60:101027, 2020.

\bibitem{Nagrani2017VoxCeleb}
A.~Nagrani, J.~S. Chung, and A.~Zisserman.
\newblock {Voxceleb: A large-scale speaker identification dataset}.
\newblock {\em arXiv}, pages 2616--2620, 2017.

\bibitem{NIST2021Eval}
NIST.
\newblock Nist evaluates face recognition software’s accuracy for flight
  boarding.
\newblock 2021.

\bibitem{Pahl2022FemaleFaces}
J.~Pahl, I.~Rieger, A.~M{\"{o}}ller, T.~Wittenberg, and U.~Schmid.
\newblock {Female, white, 27? Bias Evaluation on Data and Algorithms for Affect
  Recognition in Faces}.
\newblock In {\em ACM Fairness Accountability and Transparency (FAccT)},
  volume~22, page~15. ACM, 2022.

\bibitem{Parkhi2015Deep}
O.~M. Parkhi, A.~Vedaldi, and A.~Zisserman.
\newblock {Deep Face Recognition}.
\newblock In {\em British Machine Vision Conference}, 2015.

\bibitem{Paullada2021DataResearch}
A.~Paullada, I.~D. Raji, E.~M. Bender, E.~Denton, and A.~Hanna.
\newblock {Data and its (dis)contents: A survey of dataset development and use
  in machine learning research}, 11 2021.

\bibitem{Raji2021AboutEvaluation}
I.~D. Raji and G.~Fried.
\newblock {About Face: A Survey of Facial Recognition Evaluation}.
\newblock Technical report, 2021.

\bibitem{Rodrigues2012}
C.~Rodrigues and D.~Simões.
\newblock How can sociolinguistic data be used?
\newblock {\em Revista Diacrítica}, 27:287--308, Dec. 2012.

\bibitem{Roux2014ADatasets}
L.~Roux, J.~. Vincent, J.~Le~Roux, and E.~Vincent.
\newblock {A categorization of robust speech processing datasets}.
\newblock Technical report, 2014.

\bibitem{Sadjadi2016}
S.~O. Sadjadi, S.~Ganapathy, and J.~W. Pelecanos.
\newblock Speaker age estimation on conversational telephone speech using
  senone posterior based i-vectors.
\newblock pages 5040--5044, 3 2016.

\bibitem{Sambasivan2021EveryoneAi}
N.~Sambasivan, S.~Kapania, and H.~Highfll.
\newblock {Everyone wants to do the model work, not the data work: Data
  cascades in high-stakes ai}.
\newblock In {\em Conference on Human Factors in Computing Systems -
  Proceedings}. Association for Computing Machinery, 5 2021.

\bibitem{Seaborn2021voice}
K.~Seaborn, N.~P. Miyake, P.~Pennefather, and M.~Otake-Matsuura.
\newblock {Voice in human-agent interaction: A survey}.
\newblock {\em ACM Computing Surveys}, 54(4), 2021.

\bibitem{Singh2019Profiling}
R.~Singh.
\newblock {\em {Profiling Humans from their Voice}}.
\newblock 2019.

\bibitem{Sturim2016CorporaSystems}
D.~E. Sturim, P.~A. Torres-Carrasquillo, and J.~P. Campbell.
\newblock {Corpora for the Evaluation of Robust Speaker Recognition Systems}.
\newblock In {\em Interspeech 2016}, volume 08-12-September-2016, pages
  2776--2780, ISCA, 9 2016. ISCA.

\bibitem{Sturim2016}
D.~E. Sturim, P.~A. Torres-Carrasquillo, and J.~P. Campbell.
\newblock Corpora for the evaluation of robust speaker recognition systems.
\newblock volume 08-12-September-2016, pages 2776--2780. ISCA, 9 2016.

\bibitem{Suresh2021Framework}
H.~Suresh and J.~Guttag.
\newblock {A Framework for Understanding Sources of Harm throughout the Machine
  Learning Life Cycle}.
\newblock In {\em EAAMO '21: Equity and Access in Algorithms, Mechanisms, and
  Optimization}, 2021.

\bibitem{TDPersonalBankingVoicePolicy}
{TD Personal Banking}.
\newblock {Voice Print System Privacy Policy}.

\bibitem{ThomasHeath2016ThisBathroom}
{Thomas Heath}.
\newblock {This employee ID badge monitors and listens to you at work - except
  in the bathroom}, 2016.

\bibitem{Zheng2017robustness}
T.~F. Zheng and L.~Li.
\newblock {\em {Robustness-Related Issues in Speaker Recognition}}.
\newblock 2017.

\end{thebibliography}
}

\end{document}